# Report on the Future of Conferences

Steven Fraser, Innoxec, Santa Clara CA USA, sdfraser@acm.org

Dennis Mancl, MSWX Software Experts, Bridgewater NJ USA, dmancl@acm.org

January 9, 2023

## ABSTRACT

In 2020, virtual conferences became almost the only alternative to cancellation. Now that the pandemic is subsiding, the pros and cons of virtual conferences need to be reevaluated. In this report, we scrutinize the dynamics and economics of conferences and highlight the history of successful virtual meetings in industry. We also report on the attitudes of conference attendees from an informal survey we ran in spring 2022.

## 1. Conferences must evolve

In March 2020, catalyzed by the need for "social distancing" due to the COVID-19 pandemic, conferences, trade shows, symposia, workshops, and other "mass" meetings were canceled, postponed, or moved to virtual (online) formats for the balance of the year. Government travel restrictions and individual health concerns made in-person conferences difficult, if not impossible, to organize.

The authors wholeheartedly endorse the adoption and growth of virtual and hybrid conferences, even as the pandemic subsides. We strongly believe that conference organizers must increase conference accessibility by reducing the cost for attendees. Accessibility is improved and costs are reduced by the adoption of virtual and hybrid conference strategies. Communities that sponsor conferences need to create new and more open conference models to foster increased diversity, equity, inclusion, and accessibility while decreasing attendee cost and carbon footprint.

Pre-COVID, conferences were organized as face-to-face assemblies with participants congregating at convention centers, hotel complexes, resorts, or on company/university campuses. Attendees would meet, talk, give presentations, present papers, receive feedback, market and sell products/ideas, network, build community, and have some fun together.

Based on experiences of the past two years, participants extol the benefits of no-travel conferences. Virtual events eliminate conference-associated risks from the pandemic, reduce climate change impact, and increase accessibility for those with limited travel budgets or government travel restrictions. Others yearn for a return to face-to-face meetings, driven by a desire to return a pre-pandemic status quo with in-person networking and the attraction of interesting destinations. The authors believe that each side of the debate has merits. However, we strongly believe that virtual and hybrid public conferences will flourish, in spite of the nostalgia for pre-pandemic in-person conferences.

Our report will explore aspects of in-person, virtual, and hybrid conferences. We will examine motivations, logistics, technology, finances, and new ways of enabling interactions.

In the course of our study of conferences, the authors ran a community survey in spring 2022 to probe for opinions about the value of in-person, virtual, and hybrid conferences [1].

This report will not address the pandemic-era trend to "work from home" or the more recent debate over a "return to the office." We have already seen many recent changes in vision of the workplace of the future, including work from open-plan offices, coworking spaces, work from home, work from anywhere, and hybrid models (a mix of workplaces). A discussion of the workplace of the future is beyond the scope of this report.

This report is a general discussion of the structure of past and future conferences – and it is aimed at conference attendees, organizers, and sponsors. The historical discussion is a recap of information that is already familiar to frequent academic conference veterans. Even so, to better understand the pros and cons of virtual conferences, it is useful to revisit the benefits of traditional conferences.

## 1.1. A look back at history

Stepping back in time, many advances in technologies over the past 250 years have changed how civilization works, learns, communicates, and plays.

Commercialization of inventions such as the steam engine (accelerating manufacturing and transportation), electricity (working longer than daylight hours), telecommunications (telegraphy and telephony), aviation, and the internet have each played a role in changing society.

Inter-personal communication has made enormous progress in 25 years. Emergent technologies have enabled the virtualization of retail commerce, the evolution of print media from paper to websites, blogs, Twitter, and other digital social media platforms, and the migration from POTS (Plain Old Telephone Service) to multi-media platforms enabling text, images, voice, video, virtual reality, and augmented reality.

In the intervening years, internet technology has been adopted for general use in homes, offices, and schools. Business has somewhat reluctantly embraced "work-from-home" virtual meeting technologies as a result of the pandemic. Other virtual applications are emergent, from distance learning, tele-justice, tele-health, media live-streams, internet-gaming, to social networking. This brings us to the question of virtual conferences. Could conferences be the next domain where "virtual" achieves a more significant level of adoption above the plateau achieved during the pandemic?

Let us pose two questions before we delve into our prognostications on the future of conferences.

## 1.2. What is a conference?

For the purposes of this report, a conference is a meeting to share, discuss, and expand knowledge. Some exemplars of conferences are:

- **Academic conferences** are sponsored by professional societies, industry associations, or academic entities. The core part of an academic conference often consists of peer-reviewed research paper presentations and prepared talks; attendees come to learn, discuss, and network. Examples include the ACM/IEEE ICSE and ACM SPLASH Conferences. Key characteristics are: Low registration fees, discounts for students, all participants pay; size may range from small to large.
- **Commercial conferences** are organized by media companies like InfoQ and O'Reilly and feature tutorials, keynotes, and panels delivered by industry experts. Registration fees in comparison to academic conferences are higher and speakers are paid to present. Conference size is generally large.
- **Developer conferences** are focused on company product ecosystems, e.g., JavaOne, AWS Summit. Conference size is generally large.
- **Trade association conferences** are organized to showcase the latest products and innovations in an industry. Participants market products to increase sales, e.g., CES, Interop, Mobile World Congress. Conference size is generally large to ultra large with many tens of thousands of attendees.
- **Government/NGO conferences** initiate and continue discussions on policies and innovations with societal breadth – for example, the 2021 UN Climate Change Conference in Glasgow. Conference size varies from small to ultra large.

## 1.3. What factors influence conferences?

**Factor #1**: **Conference stakeholders.**

It is important to understand the set of stakeholders (the people who are connected directly and indirectly to conference registration and conference events) to ensure that future conferences deliver value.

Within each stakeholder category, the set of conference participants is constantly evolving. The conference community is growing more global and diverse, especially in the domain of scientific and technical conferences. Conferences have created a competitive ecosystem, but there is also a place for effective collaboration among the stakeholders.

Key stakeholders include:

- **Attendees** – individuals who will benefit directly from participation through making presentations, learning, and marketing ideas/products

- **Attendee Proxies and Sponsors** (companies, universities, or funding agencies that pay for attendees to attend a conference)
- **Conference Organizers** (individuals, companies, professional societies, NGOs, governments)
- **Conference Sponsors** – organizations that support conferences through paid sponsorships or gifts
- **Venue Sponsors** – tourism/hospitality businesses and regional development government interests

**Factor #2**: **Motivations for attendance differentiated by "personal" or "organizational" benefits:**

| | |
|---|---|
| *Personal Self-improvement* | Learning<br>Ideation<br>Problem Solving<br>Publishing<br>Networking<br>Fun |
| *Company, University, or Organization Benefit* | Learning<br>Scouting Trends<br>Recruiting<br>Marketing/Selling Products<br>Enhancing Reputation |

**Motivations for Attending Conferences**

In general, a conference is a great place to create and communicate, and there is almost always some reward in terms of individual self-improvement.

"Learning" appears twice in the table above because Individuals and organizations both may benefit from the conference "learning experience." For example, a company may send a group of employees to a conference tutorial or workshop.

**Factor #3**: **Conference finances must be economically sound.**

Revenues must balance costs. The COVID-19 pandemic has demonstrated that it is not simply a matter of "build it and they will come." Conference stakeholders (organizers, attendees, and sponsors), each have a different perspective on economics.

Organizers orchestrate conference logistics by soliciting, curating, and marketing content to put on "the show" (i.e., the conference). Organizers assume financial risks, and their "rewards" depend on the nature of the conference. In some cases, conferences are commercial ventures where the organizers hope to turn a profit (with revenue greater than the event's costs). For academic conferences, success might be measured by "breaking even" after taking into account government grants and sponsorships. Other conferences are organized by professional societies with some costs offset by society membership fees.

Conference organizers need to balance costs and revenues. Some costs are fixed (independent of the number of attendees):

| Fixed conference costs | |
|---|---|
| Insurance and legal | Registration services |
| Professional staff | Most IT services |
| Marketing/Advertising | Conference publications (proceedings, Open Access fees) |
| Security (physical and electronic) | Speaker costs |

Other costs are variable (proportional to the number of conference attendees):

| Variable conference costs | |
|---|---|
| Meeting room logistics | IT services (wi-fi, streaming, attendee support) |
| Food and beverage | Support staff |
| Hotel logistics | Conference publications |

Whether a conference is small or large, seed funding is required to cover initial planning, marketing, and deposits when booking venues or reserving IT services.

As with any personal purchase, prospective attendees should assess the advertised value of a conference before attempting to convince their "management" (corporate or academic) to "buy" a registration. Post-conference, attendees (and their proxies and sponsors) will need to assess whether they received positive value for their time and investment (registration cost and travel/living costs). This value can be demonstrated through conference reports, inspiration, key learnings, and personal experience (fun, learning, and network growth). In times of economic restraint, corporate employees might be fortunate enough to get time-off-with-pay while having to self-fund conference travel and registration. Virtual conferences, with reduced costs, can prove attractive to budget-conscious attendees.

To evaluate the delivered value of a conference, sponsors evaluate changes in sales, market opportunities, recruiting, and other business goals. However, these are often impossible to directly quantify in the short term – and corporate sponsorships frequently depend on a company's desire to do "social good" or as part of a targeted marketing campaign. Some conference sponsors are government agencies who have ongoing programs to provide funds for the support of research, regional development, and information exchanges in selected fields.

**Factor #4**: **Nostalgia for past conference experiences**

Conferences of the "Future" will change as technology, social norms, and government policy evolve. Nostalgia shouldn't be ignored, but how many times should progress be sacrificed to satisfy a core set of repeat attendees? It is useful to reflect on the following questions:

- Which parts of traditional conference experiences are most attractive to attendees and organizers?
- Will new conference formats be sufficiently engaging to attract "repeat attendees?"

Technology changes may make virtual meetings increasingly more effective. Over time, virtual meetings will feel less awkward, especially as more people use video telephony for chatting with family and friends. Government policies may constrain travel to react to worldwide crises: carbon offset requirements (global warming), quarantines (pandemics), or diplomatic issues (sanctions, armed conflicts) that may make international travel impossible. Social norms may also change – reducing the desire to travel or interact face-to-face.

Fifty years ago, if one had mentioned "games" – one would have imagined face-to-face participation on an outdoor playing field, indoor gymnasium, or across a table. Today, over 3 billion people participate in "game" experiences online – not in-person. Technology has also catalyzed everything from the evolution of retail sales from bricks-and-mortar to virtual retail shops on the web – to matchmaking.

The "Future of Conferences" will depend more than we can imagine on the evolution of technology, social norms, and government policy.

## 2. Origins of virtual conferences

Virtual meetings weren't "invented" as a result of the pandemic in the spring 2020. Virtual meetings emerged in the late 1980s for companies to reduce costs [2]. In the 2010s, learned institutions experimented with virtual conferencing to reduce their carbon footprint – for example, the Nearly Carbon-Neutral (NCN) conferences (UCSB) [3]. In April 2020, an ACM task force published a report on best practices for virtual conferences [4].

Multinational companies were already embracing virtual meeting technology in the late 20th century, which they used for business meetings and large company events. Virtual meetings helped cut travel costs and reduced "out-of-office" time. Internal company meetings made increasing use of virtual meeting technology in the late 20th century, even though public conferences remained in-person. Some large companies sponsor internal virtual forums: multi-site events to share best practices across business units [5].

As technology developed, company meetings became hybrid with a mix of in-person and virtual participation. Companies used the best communications technologies they could afford: teleconferencing in the 1980s, multi-site video rooms (ISDN-based) in late 1980s to 2000s, telepresence systems beginning in the mid-1990s, and evolving desktop video collaboration applications starting in the early 2000s. Early telepresence systems were costly to run and required: specialized rooms, high performance equipment, and special low latency high bandwidth networking. Companies welcomed the advent of desktop video collaboration applications – the earliest desktop applications (such as WebEx and Skype) were primitive, but they were cheaper, easier to use, more accessible, and scalable across enterprises.

There are social challenges associated with today's virtual meeting technology. In a hybrid meeting, with a mix of in-person and virtual, some virtual attendees feel they are "second-class" participants. Virtual attendees miss side conversations and are limited in how they can influence the course of a meeting. Virtual attendees don't always hear what was said or miss attendee body language cues. In-person interactions have a much higher "social bandwidth" than virtual interactions.

A recent ACM conference paper made this point: "[T]he most challenging asymmetry is the diverse experience between co-located and remote meeting participants. Remote participants often feel isolated, while co-located participants dominate the interaction. [6]"

For public conferences, virtual technology did not gain traction before 2020, even though there were some trials, such as ACM and IEEE's ICSE conference experiment with MBone (multicast backbone) in 1995 [7]. In the world of "virtual meeting technology," public conferences were late adopters.

Why wasn't virtual technology adopted by public conferences, even though it was being used widely in company business meetings? At the time, the authors believe, a transition to virtual public conferences would have been a disruptive change in the participation and economic models. Most decision makers (conference organizers and long-time conference attendees) were likely reluctant to make changes to successful in-person conference models. In contrast, internal company meetings are another matter: a company could easily realize significant travel savings and time savings by going virtual.

Today, progressive conference organizers should consider the need to improve conference accessibility for students, young professionals, women, and others with less influence in the conference hierarchy. A virtual conference structure might serve to expand and diversify the conference community [8].

## 3. Conferences are a business

Conferences and other in-person business meetings have been "big business." The conference business exploits the dimensions of entertainment, tourism, and wanderlust (the appeal of travel, especially to exotic locations). The economic influence of a trillion dollar conference and event industry is difficult to resist [9]. The business meeting industry advertises the charms of their meeting venues and cities to conference decision makers. In addition to the business or academic attraction of conference content, a tourist destination will attract attendees wishing to mix business and pleasure. The authors suspect that exotic conference locations raise the popularity of a conference – but it has been difficult to obtain comparative statistics.

Virtual conferences are boring in comparison to "destination" conferences.

## 4. Logistics for virtual and hybrid conferences

During the pandemic, conferences borrowed many ideas from the classroom – adapting techniques for both "all-virtual" and "hybrid" learning. Schools chose to employ virtual and hybrid to keep students and teachers safe – choosing "all-virtual" when the local infection rate was high, transitioning to a hybrid mix of in-person and virtual instruction as infection rates declined.

Hybrid enabled students to be "socially distant" in a half-full classroom, and it also helped students feel less isolated after months of virtual schooling. Teachers complained about the logistical challenges: it is very difficult to organize classroom activities that can deliver an equivalent learning experience for in-person and virtual students.

The motivation for choosing a virtual or hybrid structure is different for schools and conferences. For schools, the main motivation has been local health concerns. On the other hand, conferences are a very different context from schools. Differences for conferences (in contrast to schools) include:

- No grades
- No attendance requirements
- Sharing new information on the leading edge
- Audience a mix of experts and non-experts
- Participants from multiple time zones
- Global participants with expensive travel costs
- Condensed time schedule (few days versus school year)
- Participants attracted by well-known experts on the program
- Participants attracted by the conference's focus

## 4.1. Virtual presentations can be live or recorded

In all-virtual conferences, there are four major variations for making virtual talks and sessions available to virtual attendees:

1. **Live Program**: Program is presented as a sequence of "live" presentations; attendees may ask questions in real-time (Zoom, WebEx, …).
2. **Recorded presentations with "live" Q&A**: Each presenter pre-records their presentation which is displayed in sequence; attendees may ask questions in real-time following recorded presentation.
3. **Recorded "on-demand" presentations**: Pre-recorded presentations may be viewed by attendees in any order.
4. **Mix of Live and Recorded Presentations**: Presentations with Q&A are recorded in real time; conference attendees have two additional choices for viewing: to watch a "mirror" replay at a designated time (e.g., 4, 6, 8, 12 hours) later the same day, or "on demand" (at any time after the "live" session).

"On demand" is ideal to avoid attendee "schedule conflicts" (two or more talks scheduled at the same time). Attendees with large time zone offsets (more than 2 hours) appreciate options to view presentations at a more convenient time.

However, "on demand" does not support interactions between the presenters and the audience. Live interactions are possible only in options 1 and 2. In those options, organizers usually include a short question and answer session for each talk – and this "feedback and interaction" can be the most interesting part of a conference session. Interactive sessions need to be engaging and structured to support in-person and virtual participants equitably. Virtual/hybrid conferences can be more staffing-intensive: a standard research paper session requires multiple facilitators per session, including an in-person chair who works to keep all attendees engaged and several behind-the-scenes production assistants.

Virtual conference panels enable global participation by experts who would not otherwise attend the conference in person. For example, the authors recently organized ICSE and SPLASH virtual panels with diverse panelists from four continents unlikely to attend ICSE or SPLASH.

## 4.2. Hybrid conference options

While in-person and virtual conferences are fairly straightforward to explain, hybrid conferences have different options. A hybrid conference will include both in-person and virtual elements.

A hybrid conference could be "asynchronous," consisting of a separate in-person conference and virtual conference that are separated in time.

For example, ACM/IEEE ICSE 2022 had a virtual conference (May 10-13, 2022) and an in-person conference two weeks later (May 25-27, 2022). At "asynchronous hybrid ICSE," most of the in-person presenters were able to present their talks for both conferences, but some presenters in the virtual conference were unable to travel and give their presentations a second time at the in-person conference.

Another hybrid conference option is "synchronous," such as SPLASH 2022 (December 5-10, 2022). At "hybrid SPLASH," some presenters were in-person, others were virtual. In-person attendees and virtual attendees could view any of the talks. Last but not least, a hybrid conference could be a blend of synchronous and asynchronous events.

Below are two tables summarizing key characteristics of in-person, virtual, and hybrid conferences.

|  | **In-Person Conference** | **Virtual Conference** |
|---|---|---|
| **"Live" presentations** | Traditional in-person conference: Live in-person presenters and attendees | Virtual conference: all presenters and attendees are virtual and presentations occur in "real time" |
| **"Recorded" presentations** | In-person attendees view prerecorded conference presentations during conference | Virtual attendees view recorded conference sessions/presentations at any time during or following the conference |

**In-Person and Virtual conference characteristics**

| **Synchronous Hybrid Conference** (overlap of in-person and virtual sessions) | **Blended Synchronous and Asynchronous Hybrid Conference Options** | **Asynchronous Hybrid Conference** (no overlap of in-person and virtual sessions) |
|---|---|---|
| Attendees may be in-person or virtual: Sessions are synchronous (SPLASH 2022) | Synchronous sessions for virtual attendees plus presentations for "local" attendees at conference hubs | In-person sessions are separated in time from virtual sessions days or weeks apart (ICSE 2022) |

**Hybrid conference options**

A hybrid conference may have one or more "hubs" – a hub is a location where attendees can meet in-person, and where some presenters may deliver in-person talks. In a hybrid conference with multiple hubs in different time zones, it is preferable that virtual sessions be held at a time that is convenient for a majority of attendees. For example, if there is a North America hub and a European hub, virtual sessions could be held in the afternoon for Europe (in the morning for North America). Each hub could host a "local program" at a time most convenient to the local in-person attendees. For corporate hybrid conferences – e.g., Cisco, Qualcomm, Nortel – conference "hubs" were the regional R&D Labs plus the corporate headquarters.

# 5. Motivation for attending conferences: networking and learning

Conferences can be a place to share knowledge in a "formal" manner. In many scientific fields, publishing new research work in conference papers can be preferable to (and faster than) publishing articles in scientific journals [10].

Conferences can also be a way to share ideas informally. Conferences offer an excellent opportunity to make new connections, build networks, and to renew friendships. Conferences bring people with shared interests together. Even if they are sometimes distracted by technology (e.g., cellphones, email, Facebook, Twitter, and TikTok), attendees become more energized when they break out of their day-to-day universe of familiar faces.

The value of making new contacts is difficult to estimate. One approach might be to assess the value of increasing a "personal network" by applying Metcalfe's Law. Metcalfe's Law proclaims that the value of a computer network is proportional to the square of the number of connected users.

Extrapolating Metcalfe's Law to social networks suggests that personal network growth is more impactful for individuals with smaller personal networks.

For an individual with a 10-person network, adding 10 more contacts increases the network's value by 300%, while those with a 50-person network, adding 10 more contacts increases the network's value by only 44%.

Even if we choose a much more modest value model, such as "proportional to N·log(N)" (as suggested by Briscoe, et.al. [11]), the impact of expanding the network is still much more significant for people with small networks. Adding 10 people to a 10-person network adds 160% to the network's value, adding 10 people to a 50-person network adds 26%.

Based on the premise of network value, the individuals who might benefit most from the networking opportunities at an in-person conference are those least likely to be able to afford to travel: students, early-career professionals, and individuals who have high travel costs – or visa related challenges.

A return to an in-person-only format is very short-sighted, in terms of attempts to foster greater diversity, equity, and inclusiveness of conference participation. There is a large potential to help more people build more diverse networks, if and only if we can organize our virtual conferences to support more effective interactions.

# 6. The dynamics of building personal networks
## 6.1. Conferences have both one-way and multi-way communication

Many conferences are centered around keynotes, tutorials, and paper presentations. These talks are similar to university lectures: a one-way form of communication with limited audience interaction. Attendees also participate in interactive (multi-way) activities, such as workshops, hands-on demos, and "shows" – be they artistic, musical, multi-media, or product-oriented in nature. Other more casual settings for interaction and information sharing may include social events and informal serendipitous conversations.

## 6.2. Knowledge transfer through personal contacts

Conference participation helps disseminate and incubate knowledge that is not yet widely available. One-on-one networking is a key part of the knowledge transfer process, even in a world that has a wealth of information in electronic media.

Today, the research community depends on the materials held in digital libraries, open-source repositories, open access journals, and online forums. Static material is complemented by online education options which help the global community of software professionals upskill and expand their knowledge.

But there are pitfalls relying exclusively on non-peer reviewed knowledge sources due to a low signal-to-noise ratio (lots of noise). The internet serves up an amazing supply of scientific and technical information, practical YouTube videos, and useful social media discussions; it also hosts misinformation and conspiracy theories.

Face-to-face conference discussions make it possible to ask questions directly. Tapping into personal networks via an informal conversation or email exchange can help accelerate research.

Without personal interaction, "asking a quick question" or "having a conversation" is slowed by the constraints of distance. The personal touch matters – interaction at a conference helps build long-term relationships with experts. Conversing with an expert can be more helpful than a computer search engine inquiry.

A conference is an ideal setting for informal discussions. At home, our focus is on day-to-day job software development, research experiments, meetings, writing and reading reports, and office bureaucracy.

## 6.3. Impromptu discussions and serendipitous interactions

Information exchanges are built on discussions and interactions, and in-person meetings improve the communication. In contrast, interactions that use or apply technology (virtual collaboration tools) can be awkward. The standard rules of interpersonal interaction have not caught up to the new wave of networking tools. In general, the authors believe that face-to-face interactions still provide better support for ideation and incubation of friendships and partnerships.

As humans, we gain insight from the tone of voice, body language, and eye contact. Interpersonal interactions at conferences are not preprogrammed or prerecorded. The interactions are made much richer by:

- impromptu discussions – without previous preparation

- serendipitous encounters – random meetings with individuals one is unlikely to meet elsewhere
- serendipitous discussions – leading to valuable or interesting revelations

Conference attendees are uniquely positioned to learn from in-person impromptu discussions – ideas that are difficult to acquire in any other fashion. For example, an in-person discussion may help to understand results or spark new approaches. Serendipity often contributes to new connections and ideas. A dialog might start with introductions followed by an exchange of recent experiences: "What did you try? Did it work? Was it a key learning? A best practice? Or something to be avoided?" A short conversation can trigger an insight, inspire a concept, or initiate a collaboration.

Serendipity happens "by chance" at a conference: at sessions, during breaks, or even on conference travel. For example, Kent Beck tells the story of how he and Erich Gamma had a useful software design session on a flight from Switzerland to the United States on their journey to attend ACM's OOPSLA 1997. In three hours of discussions, they collaborated to write the first version of the popular JUnit automated unit test framework [12].

A serendipitous conversation is more than an opportunity to exchange social networking profiles or business cards – it could inspire new ideas and collaborations.

## 6.4. Do virtual conferences support serendipity?

Most virtual conferences as currently designed have limited support for one-on-one serendipitous meetings.

There are two forces at work that limit serendipitous conversations in a virtual conference: technology obstacles and social norms. The key obstacle for virtual conference technology is the inability to communicate "presence." In day-to-day conversations, we often read body language, facial expressions, and tone of voice. With virtual meeting technology, it isn't easy to read non-verbal cues across meeting participants. The camera sees only speakers' faces, video resolution is poor, and audio can be masked by background noise.

Virtual conference attendees have low expectations for interactions with other attendees. They are resigned to being a "viewer" – watching the set of "canned" presentations without interacting either with the speaker or other attendees. Virtual conferees might be multi-tasking with non-conference activities – too busy for side conversations with other conferees.

But even if current expectations are low for virtual technology, there is hope for both the present and the future. A well-designed virtual conference program is not required to follow the same structure or timeline as an in-person conference. There are many creative options for building an effective virtual conference program to catalyze more active participation.

## 6.5. Engaging virtual conference attendees

The structure of conferences must evolve to better serve virtual attendees. In-person attendees benefit more from in-person contacts, impromptu discussions, and serendipity. Virtual attendees need similar benefits – conference activities that give them a chance to be active participants, make connections with other attendees, and establish opportunities for dialog with other conferees after the conference is over.

Conference organizers should leverage experiences from social media to get participants more engaged. Conference organizers should increase participant engagement by borrowing community-building practices from social media. A simple approach to get participants engaged is to use real-time polling throughout the conference. A session chair could use a web-based tool like MentiMeter or Slido to run frequent audience polls to sustain audience engagement.

Social media can also support multiway discussions during a virtual conference. Today's social media users have opinionated exchanges with people they have never met in real life. Virtual conference participants could convene an impromptu panel discussion with participants selected via real-time social media metrics. The panelists could be the most frequent conference Twitter or Facebook posters – the posters who get the most likes or text responses to their real-time blogging of conference talks.

In many conferences, virtual participants are totally anonymous, for example, when presentations are streamed via YouTube. In other conferences, the virtual participants connect to an online conference platform, which displays a list of session attendees – and no other valuable session context, e.g., organizational affiliation, contact information, or chat links.

While some platforms encourage attendees to add a "personal profile" associated with their conference login, there is generally no incentive for participants to enter personal information, so most profiles are left blank.

To encourage participants to add to their profile, conference organizers can provide motivation through:

- registration discounts
- post-conference access to premium conference content
- forums to enable post-conference conversations among attendees with similar interests

However, it is necessary to be mindful of GDPR [13] and other privacy concerns – and to be mindful of possible abuses. Profiles are a useful way to connect attendees who have similar (or dissimilar) interests and backgrounds – while offering privacy and diversity safeguards. Different conferences will likely require different templates. To assist attendees profile information could be suggested from personal websites and public databases, with the opportunity for participants to customize as they choose.

In order to support participant interactions during the conference (via chat or web video), it isn't necessary to have the conference platform be a "completely immersive environment." Conference participants may have other ways to chat. A conference platform ought to include some impromptu text chat or user-configurable small-group video meeting capabilities. Conference attendees would then have the option to establish their own one-on-one communications (using email, Twitter, LinkedIn, Slack, Skype, or whatever) during or after conference sessions.

Some of the experts in the conference community might also volunteer to host small-group informal chat sessions – an opportunity for non-experts to meet some of the stars in the field. This is a practice that has started to become commonplace. Some recent conferences have offered virtual sessions titled "Ask Me Anything" with a keynote speaker or another notable person.

# 7. Virtual conferences – commitment to diversity, equity, and inclusion

Conference attendance costs include registration, travel, and time away from home and office. With virtual conferences, conference costs are lower because physical logistics are unnecessary (food, meeting rooms, etc.), participant travel costs are reduced (no need for transportation or hotels), and "away time" from home and office are reduced (at least proportionally to travel time). That said, some would argue that getting away from "home and office" is the attraction of attending a conference. Crista Lopes postulated that escaping to conference "destinations" at the expense of your employer or grant agency is a key motivator for some to attend a conference [14].

But in-person conferences might not be a "safe" environment for some attendees. Some people in the technical community can be outright hostile to newcomers. Some of the hostility may include racism and sexism. But a subtle hostility is elitism – rejecting academics from lower-rated universities, company participants who are not from highly-ranked research programs, and "practitioners" who just come to learn.

As noted earlier, virtual conferences are much easier for students to attend, and many people who are unable to travel appreciate the opportunity to attend conferences from home. At a town hall meeting discussion at ICSE 2022, one attendee suggested that 50 students can attend a virtual ICSE for the cost of sending one person to the in-person ICSE [15].

Hybrid conferences can draw virtual attendees who live nearby. In metropolitan areas, traffic and parking can be an enormous obstacle to attending a meeting – it may take more than an hour to navigate rush-hour traffic, especially in congested metro regions. Hybrid conferences are a way to encourage more local participants.

# 8. Financial issues

The rise of virtual conferences has created many new revenue opportunities for conference organizers, but virtual also adds new challenges. During the pandemic, organizers found it difficult to monetize virtual conferences since attendees associated conference "cost" with in-person expenses (food & beverage plus physical meeting logistics).

In-person conferences fees are usually tiered by content elements, e.g., for an entire multi-day program – or by day, by track, by workshop, or by tutorial. Different categories of attendees may be assessed different fees – for example: presenters, industry participants, students, academics, members (ACM, IEEE), etc.

For a virtual conference, the charging model can be more fine-grained. For example, if the conference presentations are organized in one-quarter, one-third, or half day blocks, there could be a charge per block. It is one way to attract attendees who are interested in a group of specialized talks, or just the keynotes. Fine-grained session charges are likely more useful for conferences with large attendance.

With conference collateral such as session recordings, organizers have the choice between making these freely available after the conference to registered conference attendees or to charge premium access fees to a wider audience. However, organizers need to be mindful of digital rights issues. Without securing blanket permissions from all attendees, only the presentations can be shared – but not the recordings of the Q&A sessions – assuming that presenter permissions are a conference participation prerequisite.

Another question is how to set pricing for virtual attendance since attendees have expectations that virtual should be cheaper than an in-person registration. This expectation is based on the assumption that there will be no cost expenditures for physical rooms, coffee breaks, lunches, meals, or social events. However, the IT and production costs may be higher for virtual conferences that go beyond simple video conferencing (Teams, WebEx, Zoom, etc.). Advanced virtual conference services may require paid production staff, which can result in a higher per-attendee cost than in-person food and beverage services.

Recruiting, training, and supporting volunteers is a significant burden on the conference organizers. There is a tradeoff: Working with volunteers can help keep costs low. In contrast, paid staff may require less training and support.

There are many more practical suggestions for virtual and hybrid conferences in the "High-Level Planning" section of the report of the ACM Task Force on virtual conferences [4].

## 9. What factors will drive the future of conferences?

The future of conferences depends on three key issues:

- Increasing costs [financial and carbon footprint] and the inconvenience of travel.
- Emerging technologies topics spawn new conferences. Research funding and commercial investments incubate new communities that need to share and innovate. New virtual conferences have a low cost of entry. This may stimulate fragmentation of existing conferences.
- New collaboration technologies will enable new conference formats. Examples may include Virtual Reality, Augmented Reality, and Gamification.

## 10. Attitudes of conference attendees: a community survey

To learn more about current attitudes about in-person and virtual conferences, the authors ran a community survey in spring 2022 [1]. Although the survey population was not a random sample of the universe of all conference attendees, it did include a range of geographies (70% of survey respondents were from North America, 23% from Europe, 7% from the rest of the world) and there were respondents from industry (77%), academia (19%), and other (4%).

The primary result of the survey: "hybrid" was the preferred conference mode.

- 54% said they preferred hybrid
- 36% preferred in-person
- 9% preferred virtual

It is possible that hybrid was preferred by many because they wanted to have the option to attend an in-person event after two years of pandemic-imposed isolation. On the other hand, hybrid might have been the top option because many respondents are still nervous about traveling – but they didn't want to bar others from being able to attend a face-to-face conference.

In the survey's text comments, respondents shared a range of opinions. Some had serious issues with virtual attendance and made a good case for choosing in-person conferences. Comments included:

- Nothing today can replace the human networking and high-intensity one-on-one networking that happens at an in-person conference

- Virtual is too much one-way broadcast
- Virtual conferences are absolutely abysmal experiences
- Virtual – I find it really difficult to pay attention

Other respondents found virtual or hybrid conferences valuable:

- I especially value the global participation of virtual conferences, this democratizes technology development and sharing
- Hybrid offers flexibility that can meet the needs and constraints of diverse potential attendees
- Hybrid is the way of the future
- Hybrid allows me to make the choice of how to attend

We asked the survey participants to list the conferences they attended. Respondents replied with an amazingly diverse set of conferences. The 331 survey respondents reported attending over 500 different conferences, everything from AAAI to ICSE to JavaOne to SPLASH to Zoomtopia. The most frequent conferences attended were software engineering conferences (such as ICSE and SPLASH), consumer products gatherings (CES), agile development conferences (sponsored by Agile Alliance or Scrum Alliance).

The survey asked respondents to rate potential obstacles for virtual and in-person conferences. The purpose of these questions was to solicit improvements for conference program design and logistics. Responses identified aspects of virtual and in-person conferences that might limit participation.

Identified challenges for virtual conferences were:

- Ineffective support for casual discussions
- Ineffective tools for interactive discussions
- Time zone issues
- Fatigue due to long virtual meetings

Identified challenges for in-person conferences were:

- Registration and travel costs are too high
- Ongoing pandemic related health risks
- Time away from family or work

The challenges for virtual conferences focused on communications and interactions – while the obstacles for in-person conferences related to economic and health issues (cost, travel, and time).

We asked how many conferences participants attended in 2021 and how many conferences they planned to attend in 2022.

- 86% of respondents reported attending at least one virtual or hybrid conference in 2021
- 79% said they would attend at least one virtual or hybrid conference in 2022

Also, the average number of "conferences attended" rose significantly per survey respondent.

- The mean number of 2021 conferences attended = 2.1
- The mean number of planned 2022 conferences = 3.5

This ratio held firm across several job roles: managers, university faculty, and industry software developers.

To increase the viability of virtual conferences, enabling casual discussions and interactive discussions requires creativity and effort by organizers and attendees. Conference organizers need to incorporate conference activities that will help the online conference community to get to know one another. Online participants will get more out of their conference experience if they would be willing to do more than just "view" talks. Questions and dialog between conference attendees help increase understanding.

Our survey suggests that in order to increase the value of in-person conferences, conference organizers need to convert at least part of their meetings to a virtual event – to attract attendees who would not normally participate because of cost and travel barriers. Organizers need to keep in mind that the effectiveness of virtual and hybrid events will vary depending on the conference content and structure.

## 11. Keep conferences simple, understand motivations of the participants

### 11.1. Practices for virtual and hybrid conferences

To reduce the complexity and increase the appeal of virtual and/or hybrid conferences, here are some practices the authors have found useful:

- Smaller conferences (fewer talks, fewer tracks, fewer days) that simplify logistics
- Shorter conference days (4 hours of conference program per day instead of 6 or 8 hours)
- Keynote talks in "the middle of day" to anchor the program
- Moderated Q&A: host-curated questions received via chat
- Easy-to-navigate conference program: with hyperlinks to program elements
- Web-based conference programs: attendee sees the schedule with automatic time zone localization for their time zone
- Support: a "help line" for technical assistance (either via chat or phone)
- Be kind to presenters: avoid scheduling 3:00 a.m. presentations
- Registration options: sell registrations "by program element" for broad spectrum conferences; also sell an "all-access" registration

Less helpful strategies ("antipatterns") include:

- "Live" anonymous chat feeds – with inappropriate, vitriolic, or profane comments
- Conference program that is difficult to navigate
- Incomplete presenter and attendee profiles

### 11.2. Motivations for attendees and organizers

Many conference activities are linked to the "motivations for conference attendance." Virtual conferences can adequately address some of them – but there are some activities that work much better at in-person conferences. [Note that ratings in these tables are the subjective opinions of the authors.]

| Attendee Motivation | Conference Program Element | In-person | Virtual |
|---|---|---|---|
| **Learning** | All program elements | +++ | ++ |
| **Ideation & Problem Solving** | Collaborative workshops | +++ | + |
| **Publishing** | Conference papers | +++ | +++ |
| **Networking** | Social networking | +++ | + |
| **Fun** | Social activities | +++ | + |

**Relative Value of Conference Program Elements (for attendee self-development)**

| Attendee/Organization Goal | Conference Program Element | In-person | Virtual |
|---|---|---|---|
| **Cost-effective Learning** | Presentations, Tutorials, Workshops | + | +++ |
| **Scout Trends** | Expert chats, Demos, Exhibits, Posters | +++ | + |
| **Social Networking** | Snacks/Lunch/Dinner/Hallway chats | +++ | + |
| **Recruiting** | Presentations,<br>Social networking | ++ | ++ |
| **Marketing Products** | Special events,<br>Sponsor receptions, Tradeshows | +++ | ++ |
| **Enhancing Reputation** | Peer reviewed papers,<br>Organization success stories,<br>Sponsor keynotes | +++ | ++ |

**Relative Effectiveness of In-person vs Virtual Conference Program Elements**

In the opinion of the authors, the list of "goals" in the left column are the principal benefits to organizations when their employees attend conferences. Recruiting and marketing are much more effective when done in person. On the other hand, cost-effective learning is a key motivation for companies to have their staff attend virtual conferences.

Conference organizers need to monitor the primary motivations of their attendees to ensure that activities will meet the needs of both repeat attendees and conference newbies.

| Conference Organizer Objective | How | In-person | Virtual |
|---|---|---|---|
| **Education / Training** | Feature "hot topics" attractive to attendees<br>Tutorials and workshops that support "collaborative and experiential learning" | ++ | ++ |
| **Maximize Revenue** | Hot topics<br>Feature "experts"<br>Targeted marketing/discounts | +++ | + |
| **Community Building** | Targeted marketing<br>Feature community experts<br>Community sponsors | ++ | + |
| **Increased Accessibility** | Global marketing<br>Sponsor attendees | + | +++ |
| **Showcase University or Company** | Promote location benefits | +++ | + |

**Relative Value of In-Person vs Virtual Conferences for Organizers**

Again, these relative assessments are the opinions of the authors. Our framework is a starting point for the reader to evaluate the most relevant tradeoffs for conference attendees, sponsors, and organizers.

A virtual conference may have a different mission than an in-person conference – and it may be judged as "successful" even if it doesn't achieve all of the goals listed above.

## 11.3. Motivations for conference presenters
Conference presenters have a wide range of motivations. Most of their goals are similar to conference attendees – especially for presenters who are members of the core community. Presenters may also attend the full conference to learn, scout tech trends, recruit, and market products. One of the most important collateral benefits of being a conference presenter is the potential for an increase in reputation as a subject matter expert.

Some non-community members (i.e., individuals who have not attended previous conferences associated with the community) may be invited to participate in a conference program as a keynote speaker or panelist. Each conference has its own guidelines for keynotes and speaker compensation. Featured speakers might include:

- Famous researchers, authors, and experts
- Celebrities: executives, entertainers, athletes, writers, and inspirational individuals

Compensation can be an issue for speakers. An invited speaker may have a mercenary interest. Speaker bureaus have a reputation for negotiating high appearance fees for famous individuals. Other invited speakers may be willing to forego direct compensation, because they view their appearance as a publicity and marketing opportunity for their company's products and services.

"Virtual speakers" – speakers who aren't required to travel – can sometimes be less expensive. Most speakers are more willing to deliver a virtual talk, because they can avoid the time and inconvenience of travel. "Virtual" can simplify scheduling, and it is especially useful for organizing panel sessions – where panelists might participate from any continent.

On the other hand, when a virtual talk is broadcast online to a large conference audience, it raises the question of "digital rights management" which if not addressed might lead to the illegal bootlegging of screen-capture recordings by conference attendees – however this can now be a challenge for in-person conferences too! Alternatively, speakers might desire larger fees for a wider distribution of their presentations – although in the age of YouTube videos and TedTalks – the world is moving slowly towards "Open Access."

## 12. Reflecting on the past, trying new things in the future

### 12.1. Conference surveys, retrospectives, and experiments

It is essential that conference organizers keep asking questions of their stakeholders – to sustain a conference's relevance. Every conference should run a post-conference survey to identify trends (year over year) and run a retrospective to learn and improve. Also, because the best practices for virtual and hybrid conferences are evolving, organizers should experiment with new approaches.

Conference organizers need to track the changing attitudes of their own community of conference attendees and presenters, just as the authors' (Fraser/Mancl) community survey in the spring of 2022 gathered opinions from people who attend a spectrum of conferences. There are many things for conference organizers to assess:

- Are attendees satisfied with the conference's virtual platform? If not – why not?
- Is the conference is serving the needs its community? For example, if the conference is intended to serve an international audience, how successful is its marketing? How effective is the conference at delivering value?
- What changes might improve accessibility and attract a diverse community?

A "retrospective" is an essential management practice for conference stakeholders to drive ongoing improvements to a conference. A retrospective is an informal meeting of conference organizers after the event to reflect on which strategies worked well or what to consider for the next event (assuming a conference series).

A survey is an essential part of the feedback process. Why collect opinions from conference attendees? The views of community members evolve over time. Organizers might believe a virtual or hybrid conference cannot be successful based on a dated pre-pandemic survey. Attitudes change. Expanding access, reaching out to an international community, and improving diversity are also increasingly important goals for conferences in the third decade of the 21st century.

### 12.2. Attendee self-assessment after a conference

The authors have attended many conferences, and we are still learning. We have personally assessed our likes and dislikes about in-person, virtual, and hybrid conferences – because we perform our own "self-assessment" after each conference experience.

For example, we have both worked for many years in a corporate culture where technical staff members would write and present short "trip reports" following any outside activities. A good conference report focuses on two things.

- What did the conference attendee learn at the conference? (short summaries of interesting presentations, ideas collected from hallway conversations)
- How well did the attendee's conference activities meet personal and corporate goals? (learning specific technology trends, recruiting new staff, or doing targeted marketing)

With a series of self-assessment reports for conferences, the value of participation becomes more evident. In the mix of in-person, virtual, and hybrid conferences, reports help us to assess the effectiveness of each type of conference participation. Although it requires daily effort during the conference, a report doesn't need to be a long narrative. A report might consist of one or two paragraphs summarizing each day's program – an outline of conference topics, good questions from the conference sessions, and a short list of "new contacts."

The authors look forward to learning more about the evolution of conferences from new surveys shared by conference organizers and informal conference reports shared by our colleagues.

## 12.2. Attendee multi-tasking at a conference

Two situations to consider unrelated to the conference focus: (1) an in-person conference where attendees are distracted for reasons directly unrelated to the conference (work or personal); or (2) a virtual conference where attendees are multi-tasking on non-conference related matters. The degree of "distraction" is likely due to an attendee not being fully engaged by the conference or not having any conference related "deliverables." For example, will they be evaluated post-conference via a trip report/presentation which requires them to remain focused – or is their personal value self-assessed?

# 13. The future?

Although "virtual" changed the conference experience in the early stages of the pandemic, the flexibility of virtual participation has had important side benefits. Virtual has led to increased conference accessibility through lower attendee travel costs (money, time, carbon, government visas), and reduced expenditures by companies, universities, and governments. One conference category where "virtual" meeting technologies are not making a significant difference currently – are trade shows. For example, the annual Consumer Technology Association's CES 2022 reported a 75% drop in attendance (40,000 attendees instead of the pre-pandemic 170,000 attendees) [16].

Areas ripe for "virtual" improvement include:

- Increased support for casual serendipitous interactions
- Support for interactive discussions for ideation
- Convenient post-event access to video and presentation content
- Sustainable revenue models for virtual events

While "online fatigue" is an issue for virtual conferences, it is not obvious whether this is more than the fatigue experienced with travel to an in-person conference. While some might argue that the lure of a conference destination mitigates travel fatigue – the authors suggest that more data and research is required to assess the comparative impact of online versus travel fatigue on conference attendees. In-person conferences appear to catalyze increased attendee interaction. In comparison, virtual conference environments, as currently implemented, seem to foster less engagement between attendees.

The questions of in-person, virtual, or hybrid conferences – and the appropriate ways to apply advances in virtual technology – need to be answered in the context of global and personal issues. The world is facing a climate crisis. Society is becoming more aware of diversity and equity issues. All of us are facing individual challenges: keeping our knowledge and skills up to date, expanding our set of personal contacts, and protecting our health by reducing unnecessary travel. Our employers and sponsors are trying to get maximum value at the lowest cost. There is no single simple answer for conference organizers, but we should all work together to try new ideas.

The authors believe that it is short-sighted and reduces accessibility if all conferences return to an in-person format. Our informal survey suggests that individuals prefer conference attendance options. While there is more work to be done to make virtual conferences truly effective and collaborative – we should all recognize that reducing travel and carbon footprints is a good thing. In our view, we all need to foster the adoption of virtual conferences – beyond the plateau achieved during the pandemic.

*"The only thing we know about the future is that it will be different."* – Peter Drucker

*"Alone we can do so little; together we can do so much."* – Helen Keller

## 14. About the authors

This report reflects the opinions and the combined 50+ years' experience with in-person, virtual, and hybrid conferences of the authors. Fraser and Mancl have participated and presented at over one hundred ACM, IEEE, Agile Alliance, and university hosted conferences and workshops – including the 2000, 2021, and 2022 virtual and hybrid ACM SPLASH, ACM/IEEE ICSE, and Agile Alliance's XP conferences.

Fraser pioneered virtual hybrid corporate forums at Nortel, Qualcomm, and Cisco Systems starting in the 1990s with ISDN based videoconferencing (25+ sites worldwide). The Nortel Design Forum (a global internal hybrid technical conference with 30+ ISDN video meeting room hubs and audio/web desktop participation) attracted up to 2,000 attendees per forum and ran through more than a dozen editions featuring peer-reviewed paper presentations, keynotes, and interactive workshops. The hybrid QTech and CTech forums at Qualcomm and Cisco used a combination of desktop video applications (e.g., WebEx) and TelePresence – with the program anchored by in-person presentations at corporate headquarters.

Mancl has been a presenter at internal technical conferences on software tools and technology at Lucent and Alcatel-Lucent beginning in 1990. He also has worldwide experience in corporate education and training, developing a wide range of software technology courses and delivering them in person and virtually using multiple generations of collaboration technology.

## ACKNOWLEDGMENTS

Thanks to our anonymous reviewers and a special thanks to Moshe Vardi, Crista Lopes, and Dave Parnas for their perspectives on conferences. We would also like to thank Robert Crawhall and Steve McConnell for feedback on a draft of this report. Lastly, we would like to thank Teresa Foster and Ellen Grove from the Agile Alliance for their support of our community survey on conference preferences that the authors ran in the spring 2022.